\documentclass[twocolumn,aps,eqsecnum,superscriptaddress,showkeys,amsmath]{revtex4}		
\usepackage{graphicx}
 
\begin{document}

\title{Probing intensity-field correlations of single-molecule surface-enhanced Raman-scattered light}  

\author{O. de los Santos-S\'anchez}
\email{octavio.desantos@gmail.com}
\affiliation{Instituto de Ciencias F\'{\i}sicas, Universidad
Nacional Aut\'onoma de M\'exico, Apdo. Postal 48-3, 
Cuernavaca, Morelos 62251, M\'exico}

\date{\today}

\begin{abstract}
In the context of the quantum-mechanical description of single-molecule surface-enhanced Raman scattering, intensity-field correlation measurements of photons emitted from a plasmonic cavity are explored, theoretically, using the technique of conditional homodyne detection. The inelastic interplay between  plasmons and vibrations of a diatomic molecule placed inside the cavity can be manifested in phase-dependent third-order fluctuations of the light recorded by the aforesaid technique, allowing us to reveal signatures of non-classicality (indicatives of squeezing) of the outgoing Raman photons.  
\end{abstract}

\keywords{Raman scattering, SERS, phase-dependent fluctuations, squeezing, plasmonics}
\maketitle

\section{Introduction} \label{sec:1}

The development of ultrasensitive instrumentation based upon plasmonic devices on the nanoscale has revived, in recent years, interest in enhancing or modifying the interaction between matter and light in the quantum regime. Investigations in the area of molecular spectroscopy, at the single-molecule level, have particularly benefited from it, wherein the spectroscopic technique known as surface-enhanced Raman scattering (SERS) has played a preponderant role \cite{ru,benz,lombardi}. It is well-known that Raman scattering of molecular species is an inelastic process in which the photon scattered by a given  molecule is at a different frequency from that of the incident photon, which, in turn, provides us with information about the vibrational energy level structure of the molecule itself. In a generic SERS configuration, the inelastic scattering from a molecule turns out to be enhanced by placing it within a highly confined field at the interface of a plasmonic cavity, which stems from the photoillumination of the metallic nanostructure giving rise to a localized surface plasmon resonance \cite{kern,nabika}. For this plasmonic light enhancement to be performed, a variety of nanostructures can be used, such as (among others) plasmonic resonators \cite{zhang,alonso,yampo,baumberg}, nanowire waveguides \cite{huang,kress,lee}, nanoantenas \cite{peyskens}, nanorods \cite{basske} and plasmonic-photonic hybrid cavities \cite{barth}; the employment of metallic plasmonic nanogaps in the vicinity of particle dimers \cite{xu,talley,zhu} is also an archetypal example. Indeed, surface-enhanced molecular spectroscopy has paved the way for exploring novel nonlinear effects and tailoring light-matter interaction scenarios \cite{takase}. \\

On the other hand, since the publication of the influential work of Hanbury-Brown and Twiss \cite{hanbury}, experimental and theoretical studies of quantum fluctuations of light have mostly been focused on measuring correlations between pairs of photodetections, i.e., the particle aspect of light. Yet  another important aspect of it is also its wave character; accordingly, studies of quantum fluctuations of light   should not only admit the possibility of scrutinizing these facets independently of each other, as usual in the study of squeezing of the variance of the electromagnetic field amplitude \cite{walls,loudon}, but should also treat both wave-particle facets jointly. Carmichael and colleagues \cite{carmichael,foster1} recently addressed this issue by proposing an extension of the standard Hanbury-Brown and Twiss intensity interferometer, which consists, essentially, in recording the evolution of the amplitude fluctuations of an electromagnetic field on the condition that a photon is detected, also referred to as conditional homodyne detection (CHD). Spectra, non-Gaussian fluctuations and non-classicality are relevant features of light that can be assessed by the intensity-field (wave-particle-type) correlation measurements  thus recorded. We briefly delineate this framework further on and a more detailed discussion about it can also be found in Refs. \cite{foster2,denisov,carmichaelreview}. \\

Herein, we shall utilize the aforementioned CHD technique with a view to exploring theoretically the emergent nonclassical properties of light emitted from a plasmonic cavity in a single-molecule SERS configuration. The light to be probed is considered to be scattered, inelastically, by a single diatomic molecule placed in the gap of a plasmonic dimer. Today's achievements of this spectroscopic technique on the subnanoscale (enabling, for instance, the identification of specific vibrational modes of the molecule \cite{zhang}, together the ultra-tight confinement of light within the cavity) have made it necessary to resort to a theoretical modeling of these systems from an entirely quantum-mechanical perspective \cite{shalabney}. To this end, we shall make use of the optomechanical description of SERS introduced in Ref. \cite{javier1} so as to represent the interplay between the molecule and the quantized plasmons in the cavity. We are particularly interested in addressing the effect of quenching or amplifying the molecular vibrations (an achievable effect when the cavity is illuminated by a laser properly detuned from its resonance \cite{vahala}) upon the non-classicality of the Raman emission captured by intensity-field correlation measurements. To our knowledge, theoretical studies of joint  particle-wave correlations  of light in the context of SERS spectroscopy have not been previously reported. So, the present work is an attempt to offer a complementary perspective to analyze phase-dependent correlations of light in the domain of Raman emission where the Stokes and anti-Stokes signals take place, information that, in the light of the current experimental capabilities of SERS, some of them quoted at the outset, might possibly be tested and/or exploited. \\

The content of this paper is structured as follows. In section \ref{sec:2}, we outline the generic molecule-plasmon system we shall be working with and the quantum-mechanical approach to its description is succinctly  reviewed \cite{javier1,javier2}. Section \ref{sec:2} is devoted to a brief description of the optical scheme intended for probing the intensity-field correlation features of our light source, which is obtained by correlating a photon detection with the field quadrature based upon the CHD framework  \cite{carmichael,foster1}. On the basis of this measurement method, signatures of time-asymmetry of quantum fluctuations revealed by the aforesaid correlations and squeezing in the frequency domain are numerically explored and discussed in section \ref{sec:3}. Finally, some concluding remarks are given in Section \ref{sec:4}.


\section{Quantum-mechanical approach to a generic SERS configuration: A brief review}\label{sec:2}

\begin{figure}[t!]
\includegraphics[width=6.5cm, height=6cm]{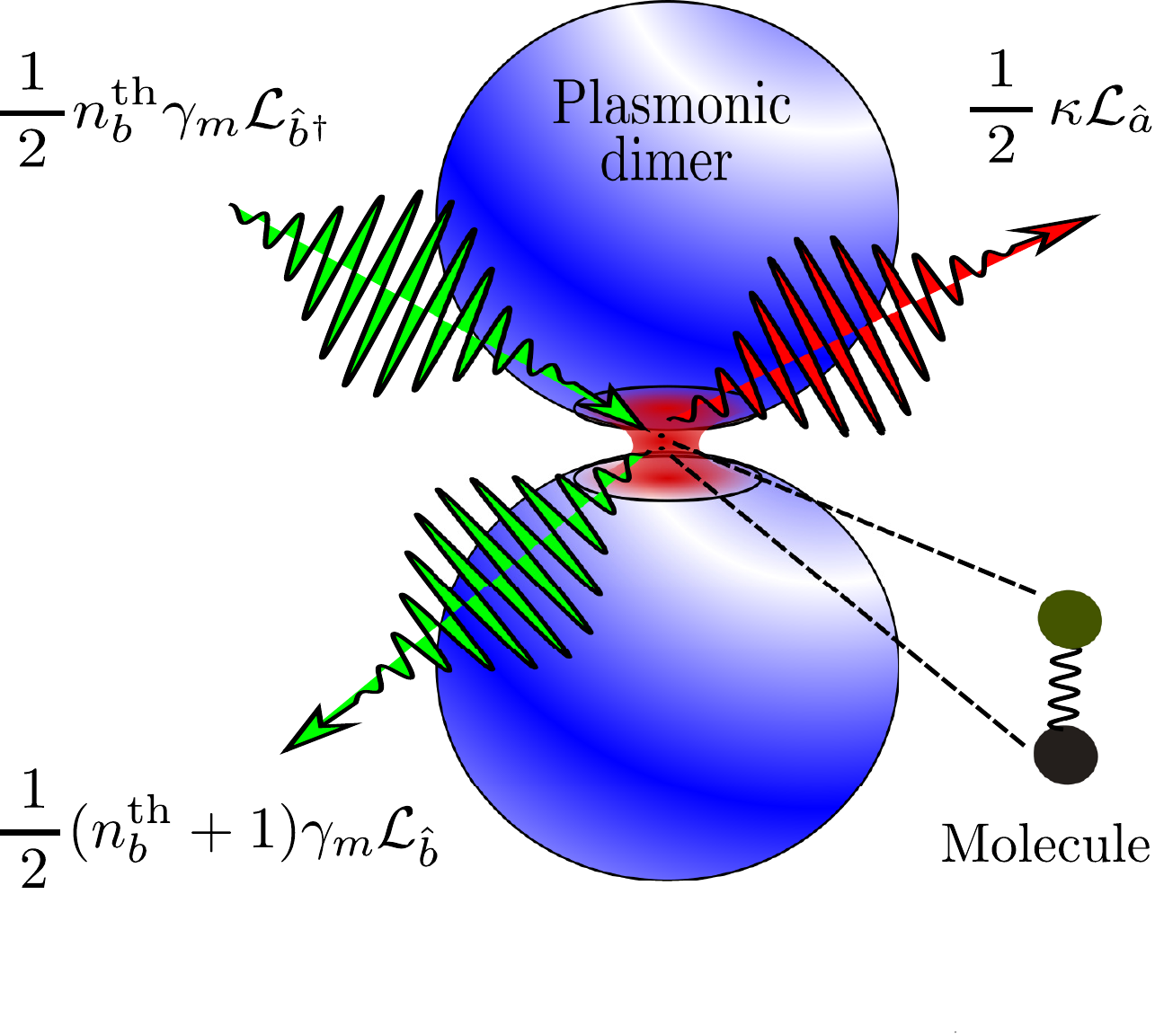} 
\caption{Sketch of a SERS setup for generating Raman-scattered light emitted from a single diatomic molecule placed in a plasmonic dimer. The open dynamics of the composite system, governed by Eq. (\ref{eq:mastereq1}), is such that the molecular vibrations are considered to undergo decay and thermal pumping processes described, respectively, by the Lindbladians $\mathcal{L}_{\hat{b}}$ and $\mathcal{L}_{\hat{b}^{\dagger}}$. The decay of the plasmic dimer, viewed as a lossy  cavity with decay rate $\kappa$, is in turn described by the Lindbladian $\mathcal{L}_{\hat{a}}$.}
\label{figure1}  
\end{figure}

Our physical system, intended for generating Raman-scattered photons, is composed of a plasmonic cavity (say, a plasmonic dimer resonator) and a single diatomic molecule placed near the surface thereof within the gap (see Fig. \ref{figure1}). 
According to the quantum-mechanical description of SERS proposed by Schmidt and collaborators  \cite{javier1} (see also Ref. \cite{javier2} for further details), the interplay between the localized plasmons in the gap and the molecular vibrations is such that the molecule, considered to be in its electronic ground state, dipolarly couples to the quantized field of the cavity via the interaction Hamiltonian $H_{I}=-\frac{1}{2} \hat{p}\cdot \hat{E}$, where $\hat{p}$ and $\hat{E}$ are, respectively, the quantized molecular polarization and electric field, the latter being evaluated at the molecule's position and explicitly given, in terms of the plasmon annihilation ($\hat{a}$) and creation ($\hat{a}^{\dagger}$) operators, by $\hat{E}=i \sqrt{\frac{\hbar \omega_{c}}{2\epsilon V_{\textrm{eff}}}}(\hat{a}-\hat{a}^{\dagger})$, where $\omega_{c}$ is the resonant frequency of the cavity, $V_{\textrm{eff}}$ its effective volume and $\epsilon$ the permittivity of the medium. In this context, the molecular polarization is also considered to be induced by the electric field (Raman-induced polarization) according to the relation $\hat{p}=\hat{\alpha}_{\nu} \hat{E}$, where $\hat{\alpha}_{\nu}$ is, in turn, the linear polarizability of the molecule. A fitting, albeit approximate, model entails regarding the molecule as a one-dimensional harmonic oscillator whose polarizability can be expressed as $\alpha_{\nu}=R_{\nu} Q_{\nu}^{0}(\hat{b}+\hat{b}^{\dagger})$; here, $\hat{b}$ and $\hat{b}^{\dagger}$ are, respectively, the standard annihilation and creation operators satisfying the commutation relation $[\hat{b},\hat{b}^{\dagger}]=1$, $R_{\nu}$ is the Raman tensor element and $Q_{\nu}$ is the zero-point amplitude of the vibrations. So, it transpires that making use of the this quantized representation for the molecular polarizability, assuming that the direction of the vibrations is optimally aligned with the field, and neglecting inessential rapidly oscillating terms at $\pm 2\omega_{c}$ frequencies, enables one to explicitly recast the interaction Hamiltonian as $H_{I}=-g\hat{a}^{\dagger}\hat{a}(\hat{b}+\hat{b}^{\dagger})$, with the real-valued coupling parameter $g=R_{\nu}Q_{\nu}^{0}\omega_{c}/(\epsilon V_{\textrm{eff}})$. This  Hamiltonian was also introduced independently by Kippenberg and colleagues \cite{kippenberg1} who suggested the analogy between the optomechanical back-action via radiation pressure force in optical cavities \cite{kippenberg2} and the modification of the energy of the plasmonic cavity by molecular vibrations in the context of molecular-plasmonic systems; this is why it is generally referred to as the optomechanical description of SERS. The whole quantum mechanical Hamiltonian governing the  coherent evolution of the isolated system is then put, after applying the rotating-wave approximation, into the form \cite{javier1, javier2}
\begin{equation}
H = \omega_{m}\hat{b}^{\dagger}\hat{b}+\omega_{c}\hat{a}^{\dagger}\hat{a}-g \hat{a}^{\dagger}\hat{a}  (\hat{b}^{\dagger}+\hat{b} ) +i\Omega(\hat{a}^{\dagger}e^{-i\omega_{l}t}-\hat{a}e^{i\omega_{l}t}),
\label{eq:ham1}
\end{equation}
where $\omega_{m}$ is the phonon frequency of the vibrational mode and where a driving laser, at frequency $\omega_{l}$, with pumping rate $\Omega$, is also added in so as to reinforce the interplay between the cavity field and the molecule. In the frame rotating at $\omega_{l}$, the above Hamiltonian can be rewritten as follows 
\begin{equation}
\tilde{H} = \omega_{m} \hat{b}^{\dagger}\hat{b}+\Delta \hat{a}^{\dagger}\hat{a}-g\hat{a}^{\dagger}\hat{a}(\hat{b}^{\dagger}+\hat{b}) +i\Omega(\hat{a}^{\dagger}-\hat{a}),
\label{eq:ham2}
\end{equation}
where $\Delta=\omega_{c}-\omega_{l}$ denotes the detuning from the cavity frequency. It is worth emphasizing that this algebraic model is restricted to the off-resonant Raman scattering regime in which a virtual state mediating the Raman transitions is in a faraway energy position from the excited electronic state. \\

To fully describe the evolution of the system, it is essential to take into account the influence of its surroundings. The incoherent effects that arise from this scenario are to be properly encapsulated in the fitting   Markovian master equation for the system density operator \cite{yampo}:
\begin{equation}
\dot{\hat{\rho}} =  i[\hat{\rho}, \tilde{H}]+\frac{\kappa}{2}\mathcal{L}_{\hat{a}}[\hat{\rho}]+\frac{(n_{b}^{\textrm{th}}+1)\gamma_{m}}{2}\mathcal{L}_{\hat{b}}[\hat{\rho}] +\frac{n_{b}^{\textrm{th}}\gamma_{m}}{2}\mathcal{L}_{\hat{b}^{\dagger}}[\hat{\rho}].
\label{eq:mastereq1}
\end{equation}
Here, the last three constituent terms are identified as the action of the Lindblad generator $\mathcal{L}$ upon the corresponding subsystem variables, that is, $\mathcal{L}_{\hat{O}}[\hat{\rho}]=2\mathcal{O} \hat{\rho} \mathcal{O}^{\dagger}-\mathcal{O}^{\dagger} \mathcal{O} \hat{\rho}-\hat{\rho} \mathcal{O}^{\dagger}\mathcal{O}$, with $\hat{O}=\hat{a}$, $\hat{b}$ or $\hat{b}^{\dagger}$, each giving rise, correspondingly, to  clear-cut processes: first, $\mathcal{L}_{\hat{a}}[\hat{\rho}]$ gives an account of the decay of photons of  the cavity at the rate $\kappa$, which, in turn, is determined by the cavity quality factor $Q=\omega_{c}/\kappa$ (the plasmonic system that takes part in our analysis is a realistic plasmonic cavity characterized by a very low quality factor: for present-day realizations, $Q \lesssim 10$); then, the terms proportional to $\mathcal{L}_{\hat{b}}[\hat{\rho}]$ and $\mathcal{L}_{\hat{b}^{\dagger}}[\hat{\rho}]$ describe, respectively, the decay and incoherent pumping of phonons of the molecular vibrations thermally excited by the environment at temperature $T$, with $\gamma_{m}$ being the mechanical decay rate and $n_{b}^{\textrm{th}}=(e^{\hbar \omega_{m}/k_{B}T}-1)^{-1}$ the effective thermal population of vibrations at frequency $\omega_{m}$. \\

\begin{figure}[h!]
\includegraphics[width=8.1cm, height=4.2cm]{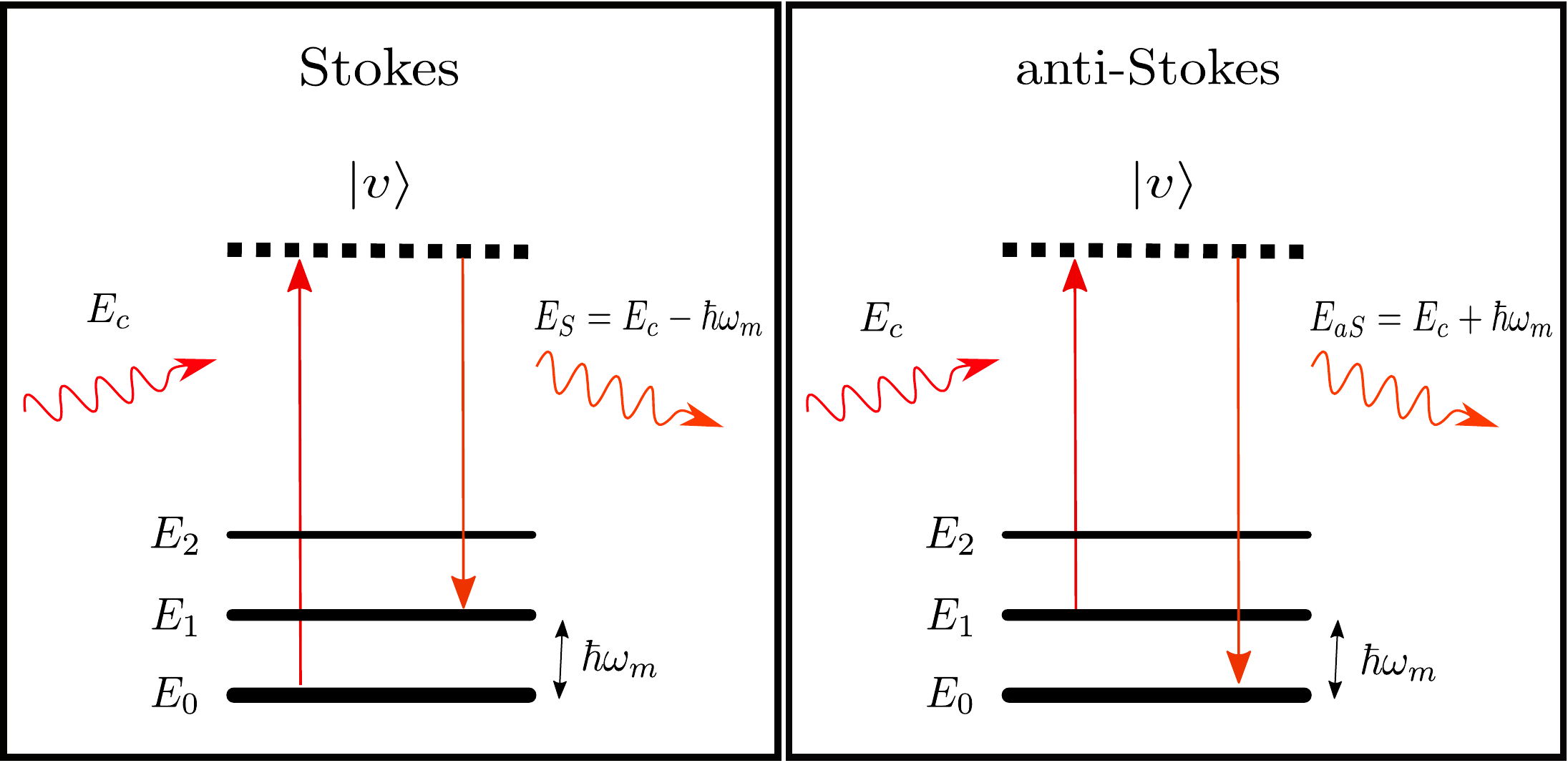}
\includegraphics[width=9.3cm, height=5.7cm]{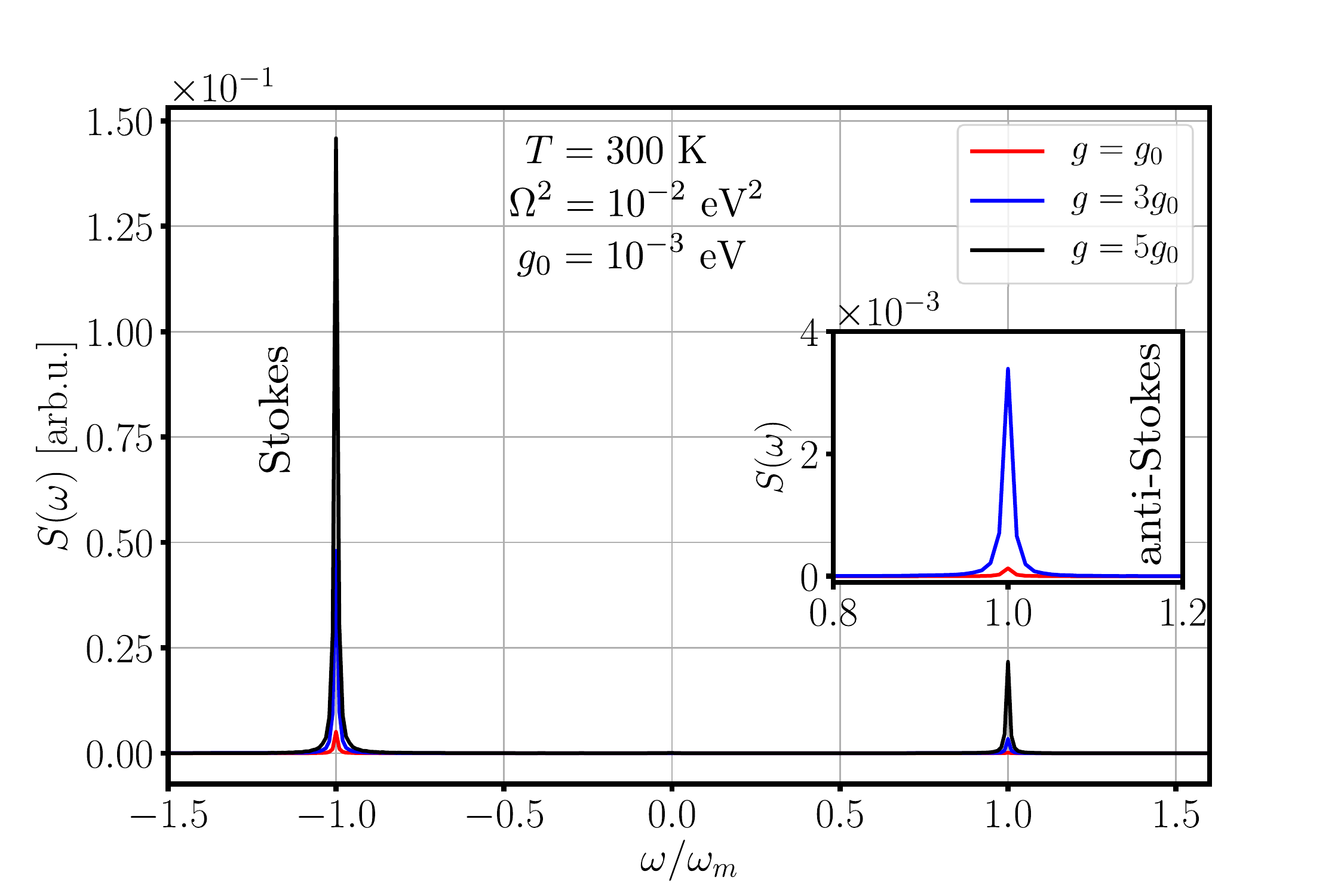} 
\caption{Upper panel: Simplified representations of the Stokes (left) and anti-Stokes (right) Raman processes, standardly understood in terms of the mediation of a virtual state $|v\rangle$. Lower panel: Emission spectra, $S(\omega)$, as functions of the scaled frequency $\omega/\omega_{m}$, of the molecule-plamon system for weak pumping, $\Omega^{2}=10^{-2}\  \textrm{eV}^{2}$, with the driving laser being tuned to the cavity ($\Delta=0$), at room temperature ($T=300$ K) and for different values of the coupling parameter as multiples of $g_{0}=1\ \textrm{meV}$: $g=g_{0}\  \textrm{(red line)}, 3g_{0}\  \textrm{(blue line)}$ and $5g_{0}$ (black line). The inset zooms in on the anti-Stokes spectral constituents for $g=g_{0}$ and $3g_{0}$. The ordinate is given in arbitrary units. }
\label{figure2}  
\end{figure}

We restrict our analysis to the spectroscopic domain where the inelastic scattering of light takes place and exhibits two archetypal components (see the upper panel of Fig. \ref{figure2}): i) the Stokes (S) spectral constituent which arises from the event in which an incident photon is converted into a phonon and a redshifted scattered photon of frequency $\omega_{S}=\omega_{c}-\omega_{m}$ (top left) and ii) the anti-Stokes (aS) constituent corresponding to the scenario where one incident photon and one existing phonon are simultaneously annihilated, thereby giving rise to a blueshifted scattered photon of frequency $\omega_{aS}=\omega_{c}+\omega_{m}$ (top right). For some increasing values of the coupling parameter $g$, the spectral fingerprint of such proceses is depicted in Fig. \ref{figure2} by computing, on the basis of solving the master equation (\ref{eq:mastereq1}), the stationary inelastic spectrum of emission $S(\omega) \propto \textrm{Re} \int_{0}^{\infty}d\tau e^{-i\omega \tau} \langle \delta \hat{a}^{\dagger}(\tau) \delta \hat{a}(0) \rangle_{ss}$, where the cavity dynamics has been split into a mean plus fluctuations in a way such that $\hat{a}(t)=\langle a \rangle_{ss}+\delta \hat{a}(t)$ (the subindex $ss$ stands for the steady-state value) and the two-time correlation function is evaluated by using the quantum regression formula \cite{carmichaelbook}. The set of parameters involved in the calculation (to be used hereafter) are within the scope of typical SERS setups \cite{yampo,javier1}: $\omega_{m}=0.1$ eV, $\gamma_{m}=1$ meV, $Q=10$,  $\omega_{c}=2.5$ eV, $\kappa=0.25$ eV, $T=300$ K, $\Omega^{2}=10^{-2}$ $\textrm{eV}^{2}$ and some multiples of  $g=1$ meV; estimations of  the molecular parameters were duly reported for the Raman activity of rhodamine 6G molecule in the supplemental material of Ref. \cite{javier1}, and BPE molecules of the type {\it trans}-1,2-bis-(4-pyridyl) ethylene, showing strong C=C stretching modes of the same order of magnitude as the previous one, also fall into the category of tractable molecular systems \cite{yampo}. So, as we can observe, both Stokes and anti-Stokes emission lines, for weak coherent pumping, reveal a conspicuous dependence on the strength of the molecule-plasmon coupling, although the anti-Stokes events are significantly less likely to take hold than the Stokes ones (as observed in the lower panel of figure \ref{figure2}). The coherent and incoherent pumping dependence of the maxima of Stokes and anti-Stokes emissions has already been addressed (see, for instance, Refs. \cite{javier1,kippenberg1,dezfouli1}) as well as intensity fluctuations of scattered photons via a frequency-filtered intensity-intensity correlation function, at zero delay, $g^{2}(0)$, observing strong photon bunching in both the weak pumping and coupling regimes \cite{javier1}. Parenthetically, the relevance of  nonclassical Stokes/anti-Stokes correlation measurements in different materials have also been remarked and examined from the viewpoint of establishing an alternative effective Hamiltonian model that incorporates explicitly such Stokes and anti-Stokes constituent fields \cite{parra}. So then, motivated by the aforementioned findings and investigations, and borrowing from the optomechanical description of SERS, we are in a position to undertake the task of analyzing the quantum fluctuations of light from a different standpoint based upon the so-called conditioned homodyne detection technique \cite{carmichael,foster1,foster2}. This theoretical background will permit us to investigate signatures concerning phase-dependent fluctuations of scattered photons through carrying out intensity-field correlation measurements.


\section{Conditional homodyne detection} \label{sec:3}

\begin{figure}[t!]
\includegraphics[width=8.5cm, height=6.5cm]{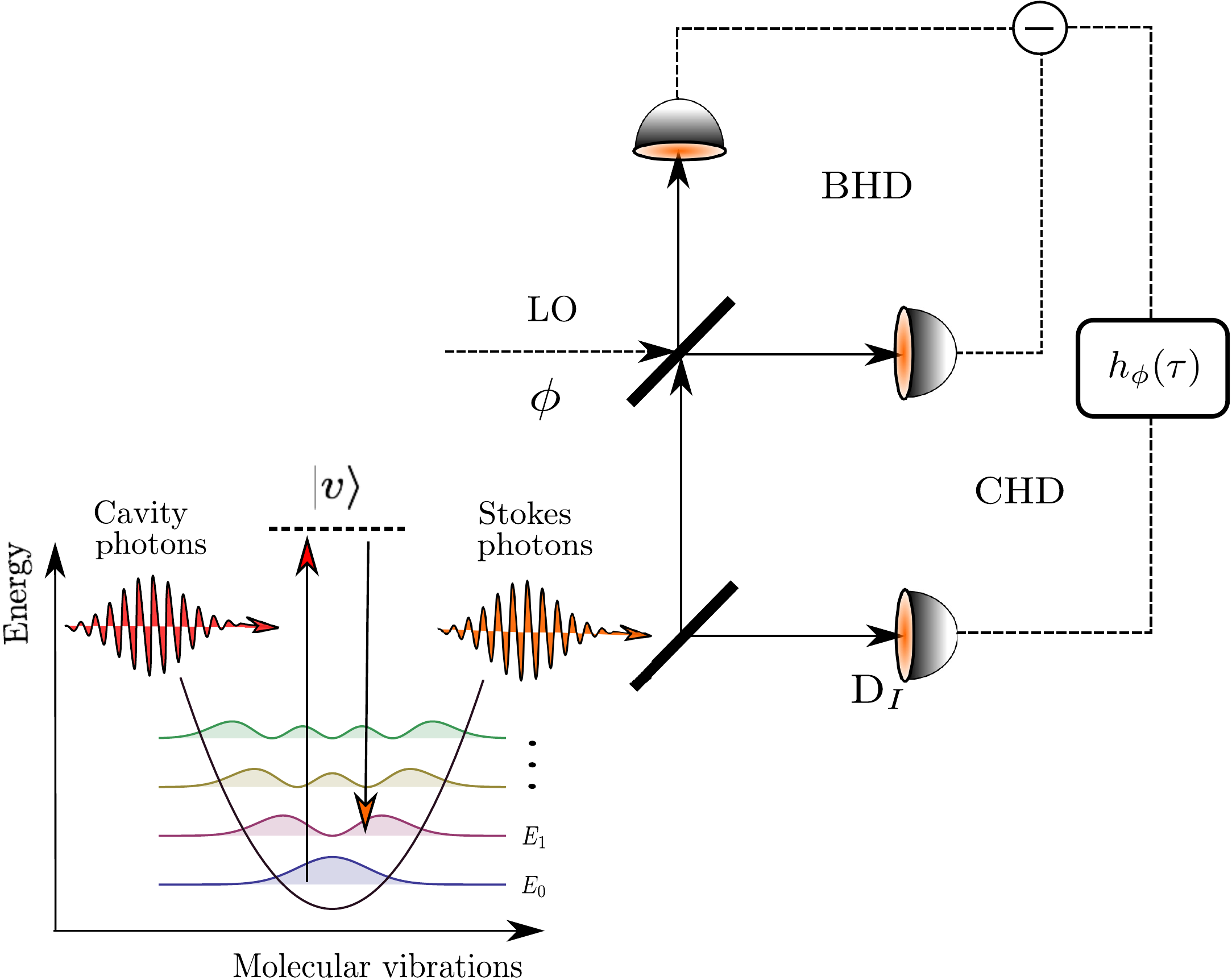} 
\caption{Top-right sketch: Optical schema for conditional homodyne detection (CHD) as if looking at the light source. The quadrature $\phi$ of the emitted field, $E_{\phi}(t+\tau)$, is measured, in the steaty state, by balanced homodyne detection (BHD) and conditioned on the photon detection, $I(t)$, at the detector $D_{I}$; LO is the local oscillator. Bottom-left sketch: A harmonic oscillator potential is used to describe the molecular vibrations leading to scattered Stokes photons.}
\label{figure3}  
\end{figure}

Originally conceived for the robust measurement of weak squeezed light in the context of a cavity QED system, the employment of conditional homodyne detection (CHD) \cite{carmichael,foster1} entails some practical advantages over standard homodyne detection techniques ascribable to its conditioning character. Pictorially sketched in the top right-hand side of Fig. \ref{figure3}, together with a basic energy configuration of the molecule-plasmon system regarded as our light source, bottom left-hand side, CHD is an optical measurement scheme in which the delayed evolution of the quadrature of the field to be probed is measured, at time $\tau$, in balanced homodyne detection, standardly composed of two detectors and a local oscillator (LO), at the exact moment of photon detection (intensity), at $\tau=0$, registered by the detector $D_{I}$. More concretely, the quantity we seek to assess, in the steady state, is algebraically embodied by the intensity-field (wave-particle-type) correlation $\propto \langle I(0) E(\tau) \rangle_{ss} $. In accord with the theoretical framework formulated by Carmichael and coworkers  \cite{carmichael}, one can scrutinize this magnitude through the normalized third-order (in the field quadrature)  correlation function:
\begin{equation}
h_{\phi}(\tau) = \frac{\langle: \hat{a}^{\dagger}(0) \hat{a}(0) \hat{a}_{\phi}(\tau) :\rangle_{ss}}{\langle \hat{a}^{\dagger}(0) \hat{a}(0) \rangle_{ss} \langle \hat{a}_{\phi}(0) \rangle_{ss}},
\label{eq:htau}
\end{equation}
where the dots ``$: \ :$" denote time and normal operator ordering and $\hat{a}_{\phi} = \frac{1}{2} (\hat{a}e^{-i\phi}+\hat{a}^{\dagger}e^{i\phi})$ is the quadrature of the quantized transmitted field, with $\phi$ being the phase between the local oscillator with respect to the averaged signal field. Incidentally, there can be circumstances in which the field amplitude of the source is such that $\langle \hat{a} \rangle_{ss} =0$, whereupon the above correlation is clearly not well defined. This difficulty can nonetheless be surmounted by merging an offset component, $E_{\textrm{off}}$, in phase with the local oscillator, with the transmitted field via an additional beam splitter \cite{carmichael}. It will not be necessary for us to resort to this adaptation in the present work. \\ 

The usefulness of CHD may be perceived as threefold. Firstly, degradation of the signal, because of  counting noise, in the measurement process is significantly reduced by the conditioning on photon detections, which makes the technique independent of finite detector efficiencies and suitable to detect very weak squeezing properties of the light source \cite{castro1}. Secondly, it allows us to find out the non-classicality of light from a different perspective, namely, the light registered through CHD  is said to be non-classical if the following  classical inequalities are violated \cite{carmichael,foster1}:
\begin{eqnarray}
0 & \le & h_{\phi}(\tau) -1 \le 1, \label{eq:ineq1}\\
|h_{\phi}(\tau)-1| & \le & |h_{\phi}(0)-1| \le 1. \label{eq:ineq2}
\end{eqnarray}
Thus, observation of non-classicality of a quadrature is not exclusively dependent upon the conventional criterion for squeezing, which states that fluctuations of such a variable must be below those of a coherent state. And thirdly, CHD is also able to reveal non-Gaussian fluctuations (nonzero odd-order correlations) of the field that can be manifested in the asymmetry displayed by the intensity-field correlation function, i.e., $h_{\phi}(-\tau)\neq h_{\phi}(\tau)$; this is also due in part to the fact that the noise properties of the field's intensity and quadrature are meant to be of different nature \cite{carmichaelreview}. This issue has been highlighted both theoretically \cite{carmichael,denisov} and experimentally \cite{foster1} in the context of cavity QED systems and subsequently, also theoretically, in resonance fluorescence of a three level atom \cite{marquina}. Still more recently, Castro-Beltr\'an {\it et al.} \cite{castro1,castro2,castro3,castro4} undertook a thorough analytical treatment of non-Gaussian fluctuations encountered in resonance fluorescence of two- and three-level atomic systems via intensity-field correlation measurements. Further context-dependent studies on the subject have also been reported, such as the light emitted from a two-level atom in an optical cavity \cite{reiner,rice}, superconducting artificial atoms \cite{wang}, fluorescence from optical transitions in $\Xi$- and V-type three-level atoms \cite{molmer1,molmer2}, and the experimental realization of the resonance fluorescence of a single trapped ${}^{138}\textrm{Ba}^{+}$ ion \cite{blatt}. \\

With the help of the quantum regression formula \cite{carmichaelbook}, let Eq. (\ref{eq:htau}) be split into positive and negative time intervals to calculate the directly measurable correlations:
\begin{equation}
h_{\phi}(\tau \ge 0)  = \frac{1}{2} \frac{\textrm{Tr} \left \{ \hat{a}(0) e^{-i\phi}e^{\mathcal{L}|\tau|} \left [\hat{a}(0)\hat{\rho}_{ss}\hat{a}^{\dagger}(0) \right ] \right \}}{\langle \hat{a}^{\dagger}(0) \hat{a}(0) \rangle_{ss} \langle \hat{a}_{\phi}(0)\rangle_{ss}}+\textrm{c.c.}, 
\label{eq:htaup}
\end{equation}
\begin{equation}
h_{\phi}(\tau \le 0)  = \frac{1}{2}  \frac{\textrm{Tr} \left \{ \hat{a}^{\dagger}(0)\hat{a}(0) e^{\mathcal{L}|\tau|} \left [\hat{a}(0)e^{-i\phi}\hat{\rho}_{ss} \right ] \right \}}{\langle \hat{a}^{\dagger}(0) \hat{a}(0) \rangle_{ss} \langle \hat{a}_{\phi}(0)\rangle_{ss}} +\textrm{c.c.}, 
\label{eq:htaum}
\end{equation}
where the superoperator $\mathcal{L}$ embraces, in toto, both the coherent and incoherent evolution of the composite system, so that Eq. (\ref{eq:mastereq1}) is encapsulated in $\dot{\hat{\rho}}=\mathcal{L}\hat{\rho}$; thus, the steady-state solution, $\hat{\rho}_{ss}$, to this equation is obtained from solving the equation $\mathcal{L} \hat{\rho}_{ss}=0$. We also stress that, for negative time intervals, the correlation given by Eq. (\ref{eq:htaum}) takes on the converse interpretation according to which the intensity measurement is conditioned on the field amplitude at $\tau=0$ \cite{castro1}. \\

The spectral fingerprint of the of CHD correlations, which can give us additional information about fluctuations and squeezing in the measurement process, can be assessed by application of the Fourier cosine transform upon them \cite{castro1,carmichaelreview}:
\begin{eqnarray}
S_{\phi}^{(\tau \ge 0)}(\omega)  & = &   4F \int_{0}^{\infty} [h_{\phi}(\tau \ge 0)-1]\cos (\omega \tau)d\tau, \label{eq:spectaup} \\
S_{\phi}^{(\tau \le 0)}(\omega) & = & 4F \int_{0}^{\infty} [h_{\phi}(|\tau|)-1] \cos(\omega \tau)d\tau,
\label{eq:spectaum}
\end{eqnarray}
where $F=2\kappa \langle \hat{a}^{\dagger}(0) \hat{a}(0) \rangle_{ss}$ is the steady-state photon flux into the correlator. In this representation, nonclassical features of light are revealed by negative values of these spectral functions. One can properly identify the origin of such features by splitting the field operator dynamics into its mean and fluctuations, i.e., $\hat{a}=\langle \hat{a} \rangle_{ss}+\delta \hat{a}$, with $\langle \delta \hat{a} \rangle_{ss}$=0. In doing so, for positive $\tau$  intervals, the intensity-field correlation (\ref{eq:htau}) is decomposed into its second- and third-order fluctuation-operator constituents, that is, $h_{\phi}(\tau) = 1+h_{\phi}^{(2)}(\tau)+h_{\phi}^{(3)}(\tau)$, where  
\begin{eqnarray}
h_{\phi}^{(2)}(\tau \ge 0) & = & \frac{2 \textrm{Re} \{ \langle \hat{a}(0)\rangle_{ss} \langle \delta a^{\dagger}(0)\delta \hat{a}_{\phi}(\tau) \rangle_{ss} \}}{\langle \hat{a}^{\dagger}(0) \hat{a}(0) \rangle_{ss}\langle \hat{a}_{\phi}(0)\rangle_{ss}}, \label{eq:fluq1}\\
h_{\phi}^{(3)}(\tau \ge 0) & = & \frac{\langle \delta \hat{a}^{\dagger}(0) \delta \hat{a}_{\phi}(\tau) \delta \hat{a}(0) \rangle_{ss}}{\langle \hat{a}^{\dagger}(0) \hat{a}(0) \rangle_{ss}\langle \hat{a}_{\phi}(0)\rangle_{ss}}, \label{eq:fluq2}
\end{eqnarray}
and $\delta \hat{a}_{\phi} = (\delta \hat{a}e^{-i\phi}+\delta \hat{a}^{\dagger}e^{i\phi})/2$. For negative intervals, one gets the correlation in the field fluctuation operator
\begin{eqnarray}
h_{\phi} (\tau \le 0)  & \equiv & 1+ h_{\phi}^{(n)}(\tau), \nonumber \\
& = & 1+\frac{ \textrm{Re}\{e^{-i \phi} \langle \delta \hat{n}(\tau) \delta \hat{a}(0) \rangle_{ss}\}}{\langle \hat{a}^{\dagger}(0) \hat{a}(0) \rangle_{ss}\langle \hat{a}_{\phi}(0)\rangle_{ss}},
\label{eq:fluqn}
\end{eqnarray}
where the superscript $(n)$ labels negative $\tau$ intervals; this quantity provides us with the same outcome as that of Eq. (\ref{eq:htaum}). Thus,  Eqs. (\ref{eq:fluq1}) and (\ref{eq:fluq2}) also enable us to split the spectral representation of fluctuations in the CHD framework, Eq. (\ref{eq:spectaup}), into $S_{\phi}^{(\tau \ge 0)}(\omega)=S_{\phi}^{(2)}(\omega)+S_{\phi}^{(3)}(\omega)$, with 
\begin{eqnarray}
S^{(2)}_{\phi}(\omega) & = & 4F \int_{0}^{\infty} h^{(2)}_{\phi}(\tau \ge 0) \cos(\omega \tau) d\tau, \label{eq:s2} \\
S^{(3)}_{\phi}(\omega) & = & 4F \int_{0}^{\infty} h^{(3)}_{\phi}(\tau \ge 0) \cos(\omega \tau) d\tau. \label{eq:s3}
\end{eqnarray}
Information about squeezed light, for instance, may be extracted from the second-order spectrum, $S_{\phi}^{(2)}(\omega)$, by virtue of the formerly demonstrated relationship $S_{\phi}^{\textrm{sq}} (\omega)=\eta S_{\phi}^{(2)}(\omega)$ \cite{carmichael,foster1}, where $S_{\phi}^{\textrm{sq}} (\omega)$ is the so-called {\it spectrum of squeezing} \cite{collet,rice2} and $\eta$ represents the the combined collection and detection efficiency; we emphasize that the conditioning character of CHD and normalization of the corresponding correlation makes the assessment of squeezing, represented by negative values of $S_{\phi}^{(2)}(\omega)$, independent of the $\eta$ factor. Beyond squeezing, the third-order fluctuations, reflected in the spectrum $S_{\phi}^{(3)}(\omega)$, would permit us to measure the extent to which non-Gaussian fluctuations, brought about nonlinearities of the light-matter interaction process, come into play. \\

So, with the foregoing theoretical tools in hand, we now proceed to the discussion of some results concerning the outcome of what we separately refer to as color-blind and frequency-filtered intensity-field correlation measurements; the latter being succinctly put forward and carried out by duly adapting the original algebraic model of the system for the CHD technique to be able to discern the frequency of the outgoing Stokes  photons. 


\section{Results and discussion} \label{sec:4}

\begin{figure}[t!]
\includegraphics[width=9.5cm, height=6.3cm]{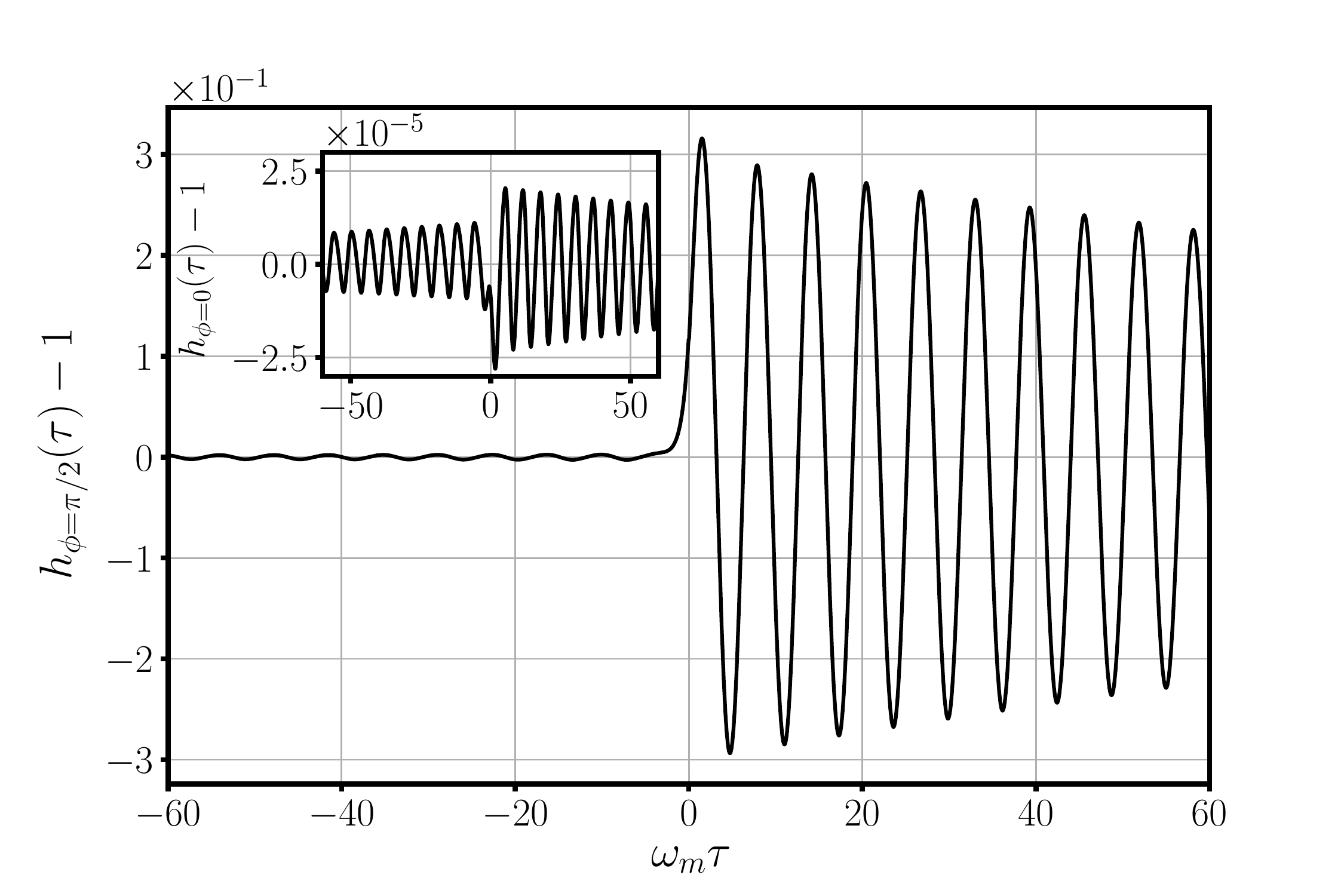} 
\includegraphics[width=9.5cm, height=6.3cm]{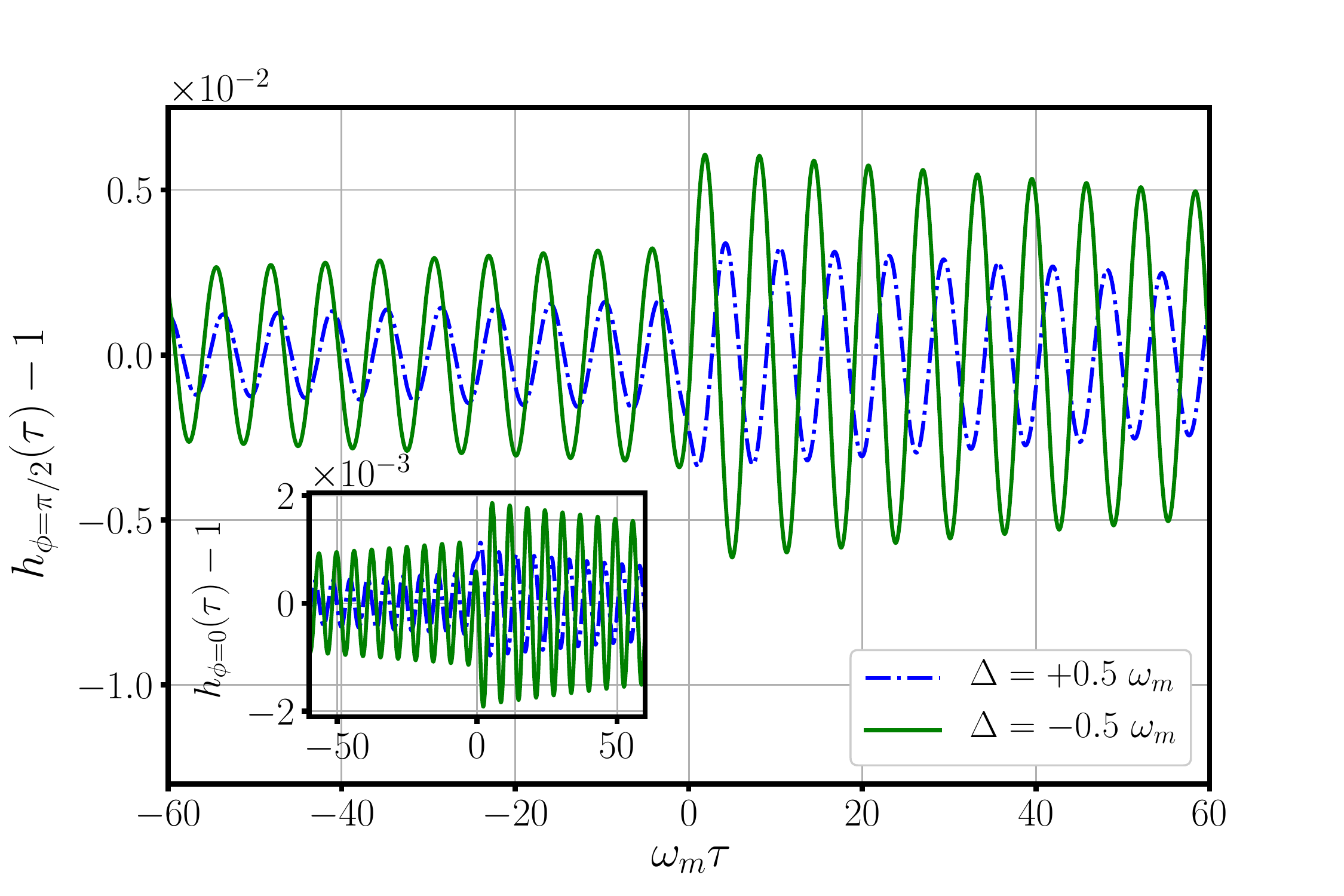} 
\caption{Intensity-field correlations, Eqs. (\ref{eq:htaup}) and Eqs. (\ref{eq:htaum}), versus the scaled time $\omega_{m} \tau$, for the $\phi=0$ (insets) and $\phi=\pi/2$ (main panels) quadratures, calculated for the moderate pumping $\Omega=1.5 \omega_{m}$. In the upper panel the laser is tuned to the cavity ($\Delta=0$), whereas the lower panel displays representative correlations in the heating ($\Delta < 0$) and cooling ($\Delta >0$) regimes. In all cases $g=5\ \textrm{meV}$ and $T=300$ K.}
\label{figure4}  
\end{figure}

In our numerical calculations we make use of the Python toolbox QuTip \cite{nori} to solve the master equation (\ref{eq:mastereq1}) and its extended version to be outlined in this section. We also restrict our calculations to weak and moderate pumping rates, $\Omega=\{0.3, 1.5\}\omega_{m}$, detuning $\Delta=\{0,\pm 0.5\} \omega_{m}$, with $\omega_{m}=0.1$ eV, and letting strength of the coupling be $g =5$ meV; the remaining parameters are taken to be fixed: $\gamma_{m}=1$ meV, $\kappa=0.25$ eV; these are realizable and of the same order as those indicated in the introduction of the emission spectra in section \ref{sec:2}. At this point, as far as the pumping coefficient $\Omega$ is concerned, it is important to underline that although we are approaching the regime where the aforementioned values of it and the cavity frequency may become comparable, such values are expected to be sufficiently moderate so that the counter-rotating terms $i\Omega(\hat{a}e^{-i\omega_{l}t}+\hat{a}^{\dagger}e^{\omega_{l}t})$, ruled out in the original Hamiltonian (\ref{eq:ham1}), do not have a significant influence upon the dynamics of the system; this issue has also been highlighted within the supporting information of Ref. \cite{javier1}.

\subsection{Color-blind intensity-field correlation measurements}

The time-dependent measurable formulae outlined above, specifically, Eqs. (\ref{eq:htaup}) and (\ref{eq:htaum}), will permit us to explore the outcome of color-blind photon correlations, so called because the present treatment assumes broadband detectors that do not differentiate between correlating photons of specific Stokes or anti-Stokes frequencies in the process; i.e., the formulae are not selective in frequency, albeit such an information may be encapsulated in their Fourier transform as stated in the previous section. Even so, with these tools, non-classical properties of the emitted field can be picked up by CHD within the whole Raman emission domain of interest to us. \\

Figure \ref{figure4} shows the outcome of intensity-field correlations, as functions of the scaled time $\omega_{m}\tau$, within a short $\tau$ interval, regarding the in-phase ($\phi=0$, insets) and out-of-phase ($\phi=\pi/2$, main panels) quadratures of the field and moderate pumping, $\Omega=1.5 \omega_{m}$. The upper panel of the figure displays, in the first instance, the case in which the laser is tuned to the cavity, $\Delta =0$, where we see that the behavior of the correlation exhibits a significant time asymmetry and that both quadratures violate the lower bound of the first classical inequality, Eq. (\ref{eq:ineq1}), and Eq. (\ref{eq:ineq2}) at certain time intervals, thus revealing some degree of non-classicality of the field. Of course, the non-classical character of the out-of-phase quadrature, for this particular case, turns out to be far more conspicuous than that of the in-phase component by several orders of magnitude (see the inset). A similar behavior was found for increasing values of the coupling parameter $g$, ranging from $1$ to $5$ meV; for brevity, we do not show any plot in this respect. In the case of non-zero detuning, it is a well-known fact that the strength of the vibrations of the molecule can be augmented or decreased depending on whether the plasmonic cavity is illuminated with a blue ($\Delta < 0$) or red ($\Delta > 0$) detuned laser, respectively, thus leading, correspondingly, to the heating or cooling of the molecule. So, it is pertinent to scrutinize the imprint of these two states of the molecule on the outcome of the correlations, as exemplified in the lower panel of Fig. \ref{figure4} for the particular values of $\Delta=\pm 0.5 \omega_{m}$. Firstly, one can see in the inset  that the detuning itself fosters the increment of the amplitude of correlations associated to the in-phase quadrature; two orders of magnitude in comparison with that of the laser tuned to the cavity frequency (inset of the upper  panel), whereas the correlation concerning the out-of-phase quadrature (main panel) is found to diminish in amplitude in marked contrast to the zero-detuning case shown in the main upper panel. Secondly, the act of heating the molecule, $\Delta =-0.5 \omega_{m}$, say, also reinforces the correlation function associated with both quadratures as we can observe by making the comparison with the cooling case, $\Delta =+0.5\omega_{m}$.\\

\begin{figure}[t!]
\includegraphics[width=9.5cm, height=6.3cm]{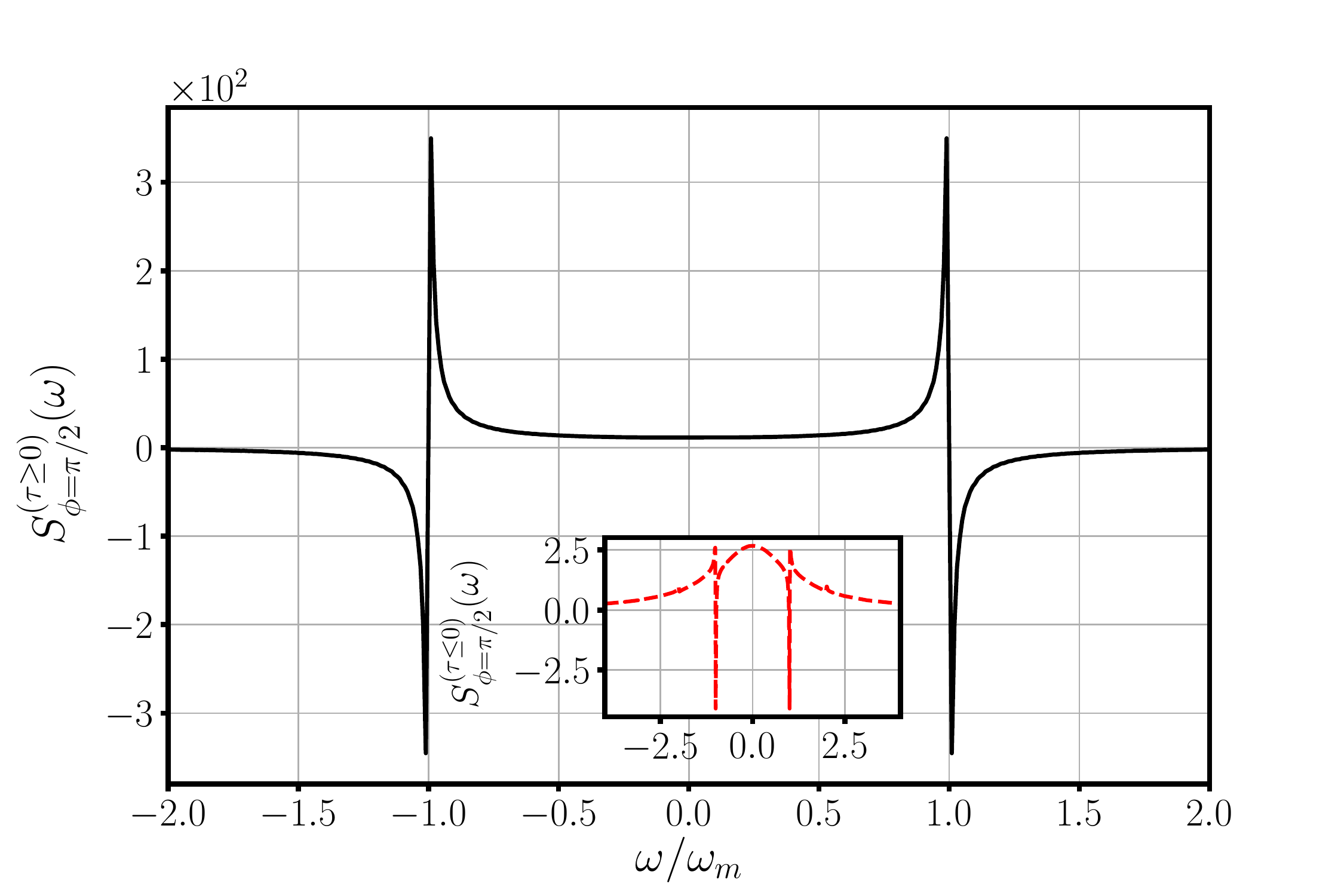} 
\includegraphics[width=9.5cm, height=6.3cm]{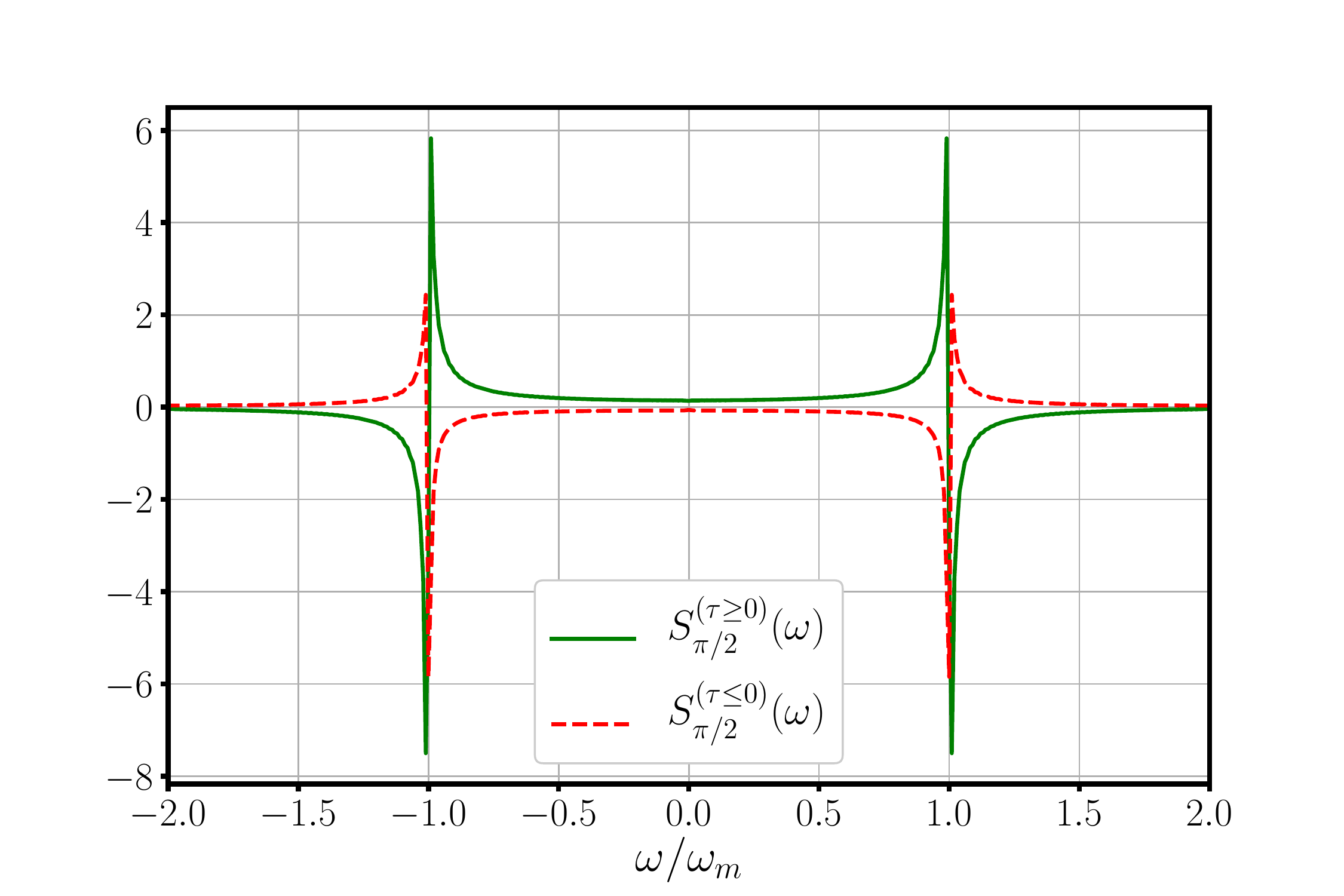} 
\caption{Spectra of the intensity-field correlation, Eqs. (\ref{eq:spectaup}), continuous line, and (\ref{eq:spectaum}), dashed line, for the $\phi=\pi/2$ quadrature and moderate pumping $\Omega=1.5\omega_{m}$. Upper and lower panels show, respectively, the spectral outcome for the zero- and negative-detuning cases: $\Delta=0$ and $\Delta=-0.5\omega_{m}$. The remaining parameters are: $g=5$ meV and $T=300$ K.}
\label{figure5}  
\end{figure}

\begin{figure}[t!]
\includegraphics[width=9.5cm, height=6.3cm]{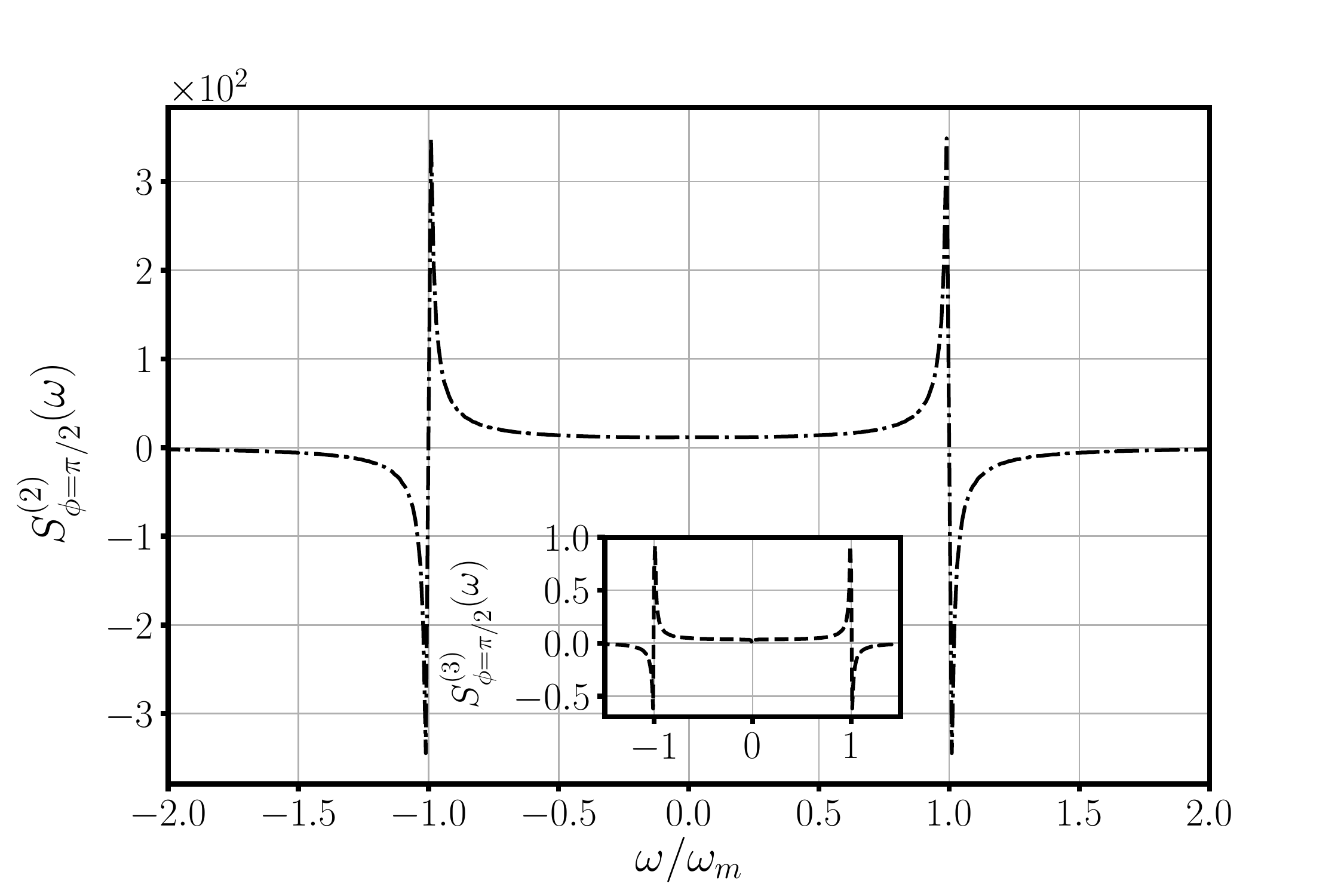} 
\includegraphics[width=9.5cm, height=6.3cm]{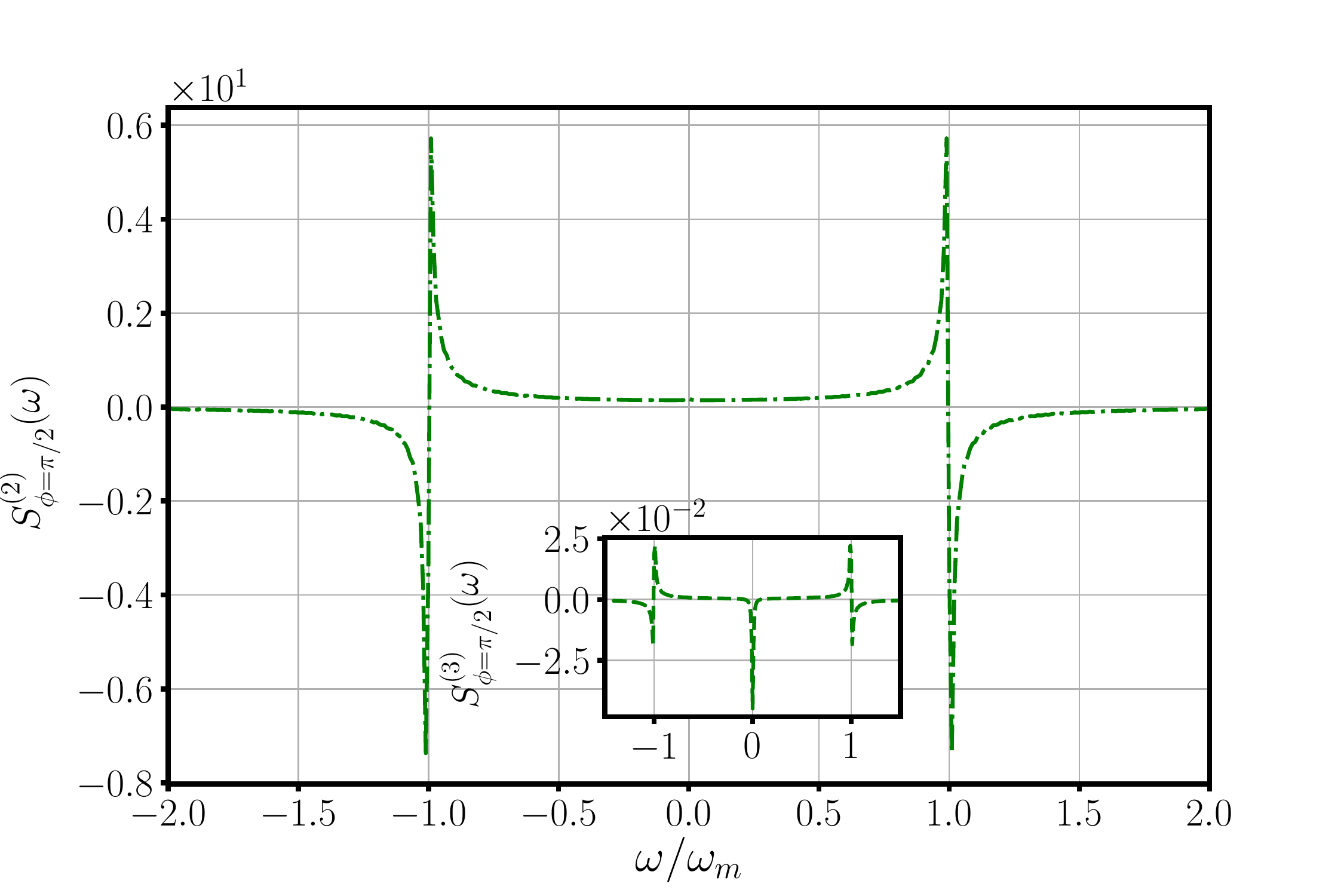}  
\caption{Decomposition of the spectrum $S_{\phi=\pi/2}^{(\tau \ge 0)}$, shown in Fig. \ref{figure5}, into its second- and third-order constituents, $S_{\phi=\pi/2}^{(2)}$ (main panels) and $S_{\phi=\pi/2}^{(3)}$ (insets), respectively. Upper and lower panels show, respectively, the spectral outcome for the zero- and negative-detuning cases: $\Delta=0$ (black line) and $\Delta=-0.5\omega_{m}$ (green line). The remaining parameters are: $g=5$ meV and $T=300$ K.}
\label{figure6}  
\end{figure}

Figure \ref{figure5} illustrates the spectral profile of some of the intensity-field correlation outcomes  described above via their Fourier cosine transform. Specifically, the upper panel of the figure displays the spectra $S^{(\tau \ge 0)}_{\phi=\pi/2}(\omega)$, main panel, and $S^{(\tau \le 0)}_{\phi=\pi/2}(\omega)$, inset, of the correlation shown in Fig. \ref{figure4}, upper panel, corresponding to the out-of-phase quadrature. For $\tau \ge 0$, the spectrum is composed of two distinctive dispersive-like profiles located at $\omega = \pm  \omega_{m}$, each attaining large negative values in a region around these anti-Stokes/Stokes positions, whereupon the light Raman scattered is confirmed as being highly non-classical. Even though the spectral function is postive over a wider frequency range concerning the $\tau \le 0$ case, see the inset, a significant reduction of fluctuations takes place also at $\omega=\pm \omega_{m}$, superimposed to what seems to be the cavity resonance contribution (wider spectral profile). The lower panel, on the other hand, exhibits the spectra of the correlation that corresponds to the particular case of $\Delta=-0.5 \omega_{m}$ displayed in the main bottom panel of Fig. \ref{figure4} associated with the in-phase quadrature. For positive $\tau$ intervals, the spectral profile (continuous line) turns out to be quite similar to the previous case of zero-detuning, whereas the spectrum $S^{(\tau \le 0)}_{\phi=\pi/2}(\omega)$ (dashed line) has inverted dispersive shapes also with conspicuous negative features near the Stokes and anti-Stokes frequencies. The origin of the aforementioned non-classical features is pictorially uncovered in Fig. \ref{figure6}, where the spectrum $S_{\phi=\pi/2}^{(\tau \ge 0)}$ is decomposed into its second- and third-order spectral constituents, $S_{\phi=\pi/2}^{(2)}$ (dot-dashed line) and $S_{\phi=\pi/2}^{(3)}$ (dashed line), respectively, for $\Delta = 0$ (top panel) and $\Delta=-0.5\omega_{m}$ (bottom panel). It transpires that the main contribution to the spectral profile and non-classicality is almost due entirely to the second-order spectral function directly associated to the spectrum of squeezing; a barely noticeable third-order contribution to the whole spectrum is understood as due to the use of a moderate pump laser (see the insets). \\

\subsection{Quadrature variances}
A customary approach to assessing squeezing entails employing the variance in a quadrature
\begin{equation}
V_{\phi} = \langle : (\delta \hat{a}_{\phi})^{2}: \rangle = \textrm{Re} \left \{ e^{i\phi} \langle \delta \hat{a}^{\dagger} \delta \hat{a}_{\phi} \rangle \right \},
\label{eq:variance}
\end{equation}
which takes on negative values for squeezed fluctuations. From the standpoint of CHD, signatures of nonclassical light can also be signaled by integrating the spectra (\ref{eq:s2}) and (\ref{eq:s3}), namely, $\int_{-\infty}^{\infty}S_{\phi}^{(2)}(\omega)d\omega=4\pi F h_{\phi}^{(2)}(0)$, $\int_{-\infty}^{\infty}S_{\phi}^{(3)}(\omega)d\omega=4\pi F h_{\phi}^{(3)}(0)$ and (\ref{eq:spectaum}), $\int_{-\infty}^{\infty}S_{\phi}^{(\tau \le 0)}d\omega=4\pi F h_{\phi}^{(n)}(0)$, obtained, correspondingly, from the positive- and negative-time interval parts of the intensity-field correlation. Given the foregoing, it suffices for our purposes to take quantities  proportional to the unnormalized correlations in (\ref{eq:fluq1}), (\ref{eq:fluq2}) and (\ref{eq:fluqn}) at $\tau=0$, i.e.,
\begin{eqnarray}
H_{\phi}^{(2)} & = & 2 \textrm{Re} \{ \langle \hat{a} \rangle \langle \delta a^{\dagger} \delta \hat{a}_{\phi} \rangle \}, \label{eq:H20}\\
H_{\phi}^{(3)} & = & \langle \delta \hat{a}^{\dagger} \delta \hat{a}_{\phi} \delta \hat{a} \rangle, \label{eq:H30} \\
H_{\phi}^{(n)} & = & \textrm{Re}\{e^{-i \phi} \langle \delta \hat{n} \delta \hat{a} \rangle \}, \label{eq:Hn}
\end{eqnarray}
where, for succinctness, arguments following the operators were omitted and the steady-state correlations, $\langle \cdots \rangle_{ss} \to \langle \cdots \rangle$, are given by the following correspondences 
\begin{eqnarray*}
\langle \delta \hat{a}^{\dagger} \delta \hat{a} \rangle & = & \langle \hat{n} \rangle-|\langle \hat{a} \rangle|^{2}, \\
\langle (\delta \hat{a}^{\dagger})^{2} \rangle & = & \langle \hat{a}^{\dagger 2} \rangle-\langle \hat{a}^{\dagger} \rangle^{2}, \\
\langle \delta \hat{n} \delta \hat{a} \rangle & = & \langle \hat{n} \hat{a} \rangle-\langle \hat{n} \rangle \langle \hat{a} \rangle, \\
\langle \delta \hat{a}^{\dagger} \delta \hat{a}^{\dagger} \delta \hat{a} \rangle & = & \langle \hat{a}^{\dagger} \hat{n} \rangle-2\langle \hat{a}^{\dagger} \rangle \langle \hat{n} \rangle -\langle \hat{a}\rangle \left (\langle \hat{a}^{\dagger 2} \rangle-2\langle \hat{a}^{\dagger} \rangle^{2}\right ).
\end{eqnarray*}

\begin{figure}[t!]
\includegraphics[width=9.5cm, height=6.3cm]{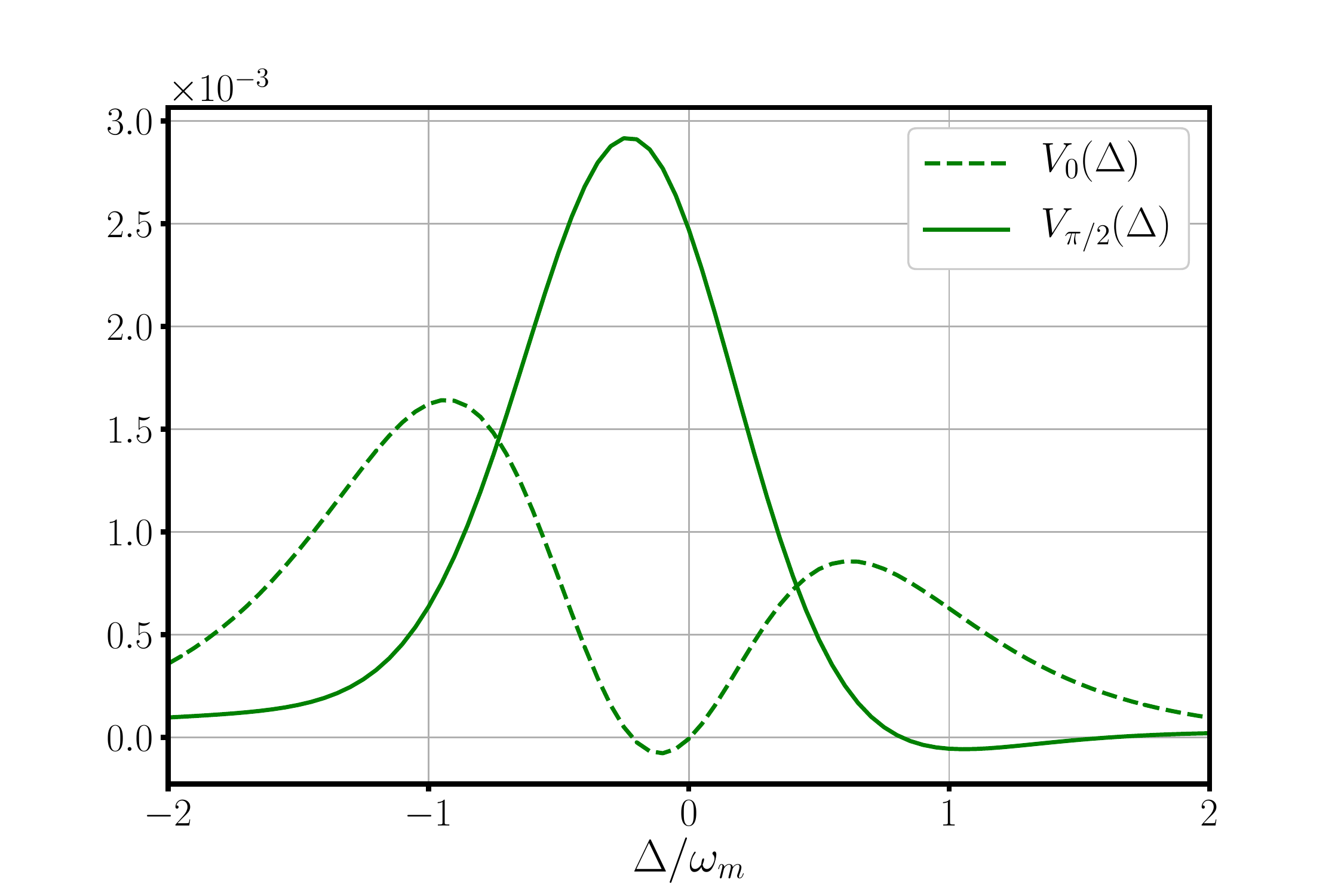} 
\caption{Variance, Eq. (\ref{eq:variance}), as a function of the scaled detuning $\Delta/\omega_{m}$, for $\phi=0$ (dashed line) and $\pi/2$ (continuous line). The parameters are: $\Omega=1.5\omega_{m}$,  eV, $g=5$ meV, and $T=300$ K.}
\label{figure7}  
\end{figure}

As a function of the detuning, we see in Fig. \ref{figure7} that, in general, the variance (\ref{eq:variance}) itself does not manifest squeezing for both quadratures, except for very narrow ranges of detuning, approximately around $-0.5 <\Delta/\omega_{m} <0$, for $V_{\phi=0}$, and within the vicinity of $\Delta=\omega_{m}$, for $V_{\phi=\pi/2}$. \\

\begin{figure}[t!]
\includegraphics[width=9.5cm, height=9.0cm]{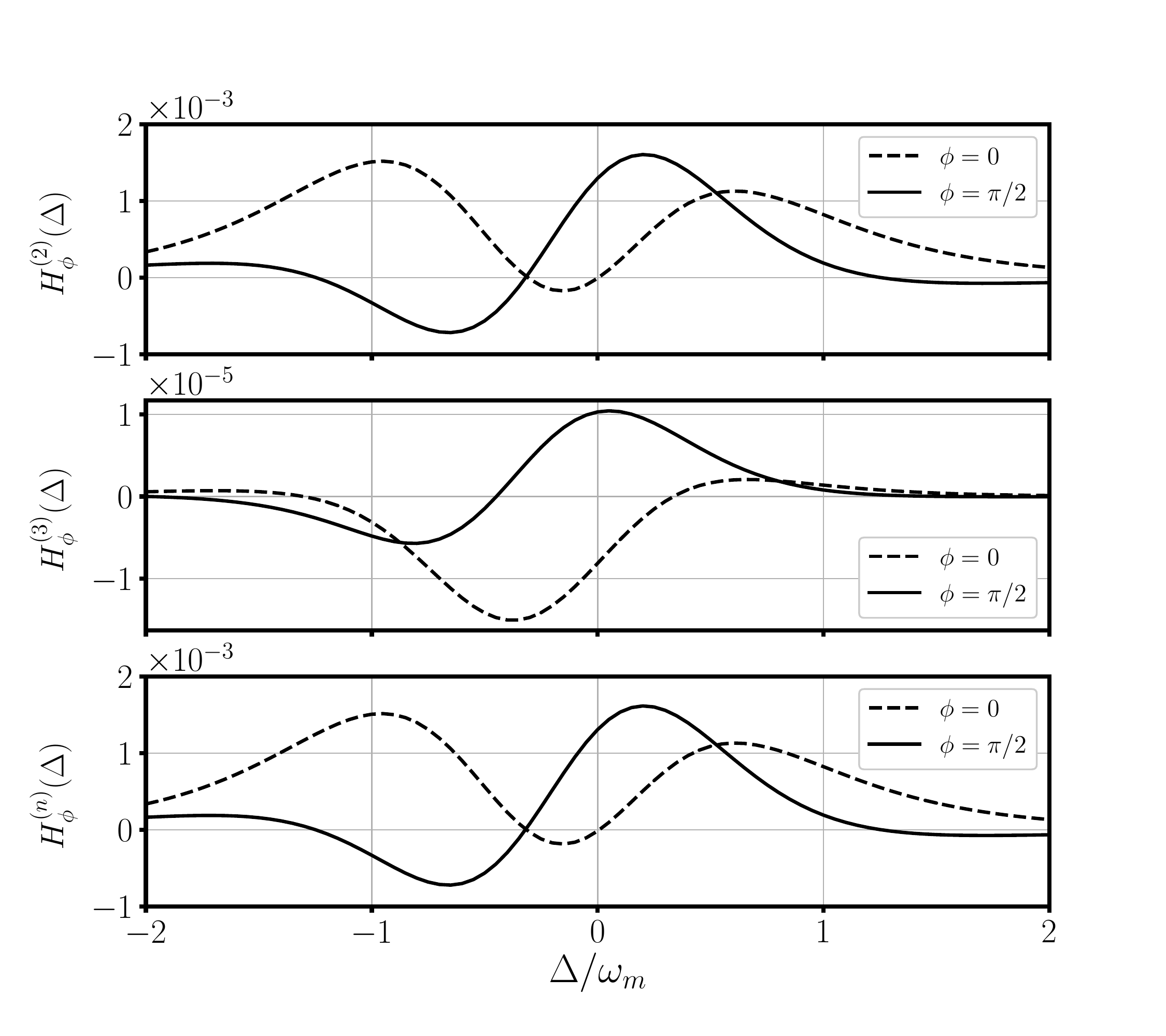} 
\caption{Noise from CHD calculated through Eqs. (\ref{eq:H20}), (\ref{eq:H30}) and (\ref{eq:Hn}), from top to bottom panels, versus the scaled detuning $\Delta/\omega_{m}$, for $\phi=0$ (dashed line) and $\pi/2$ (continuous line). The parameters are the same as in Fig. \ref{figure7}.}
\label{figure8}  
\end{figure}

In similarity to the variance, one can see in Fig. \ref{figure8} a complementary perspective by considering, separately, the noise stemming from the positive and the negative time intervals of the  correlations, and also by splitting the second- and third-order constituents of the former case, as indicated from the top to the bottom panels in the figure. The second-order term, $H_{\phi}^{(2)}(0)$, is found to be the dominating source of noise, exhibiting, in the meanwhile, negativities within certain intervals of negative detuning regarding the out-of-phase quadrature, $\phi=\pi/2$. A much lower proportion of noise (two orders of magnitude) is originated from the third-order term, $H_{\phi}^{(3)}(0)$, as seen by comparison with the previous case; some negativities are manifested mostly in the in-phase quadrature, $\phi=0$. Owing to the smallness of this term, the behavior of $H^{(N)}(0)$, see the bottom panel, is essentially identical to that of $H_{\phi}^{(2)}(0)$; at $\tau=0$, $H_{\phi}^{(n)}(0)=H_{\phi}^{(2)}(0)+H_{\phi}^{(3)}(0)$, whence it follows that $H_{\phi}^{(n)}(0) \approx H_{\phi}^{(2)}(0)$. The corresponding panel shows the result of the direct calculation of $H_{\phi}^{(n)}$ from Eq. (\ref{eq:Hn}). 


\subsection{A first glance at frequency-filtered correlations}

In section \ref{sec:3}, we presented the essential workings of the CHD scheme with which we have so far revealed signatures of phase-dependent fluctuations in a measurement process that, as such, has been indiscriminately sensible to the ``color" (frequency) of the correlated photons that our Raman signal is comprised of. Although this piece of information is valuable in itself, one could, in principle, go further in an attempt to be capable of discerning the frequency of the constituent Raman photons to be correlated. Indeed, tackling the problem of computing frequency-filtered intensity-field correlations in the context of CHD is not a simple endeavor, so that the spirit of the present subsection will be essentially algebraically exploratory. At first sight, to deal with the problem of mathematically differentiating between Stokes and anti-Stokes photons in our calculations, one would be tempted to posit an intensity-field correlation, echoing Eq. (\ref{eq:htau}), of the form: 
\begin{equation}
\tilde{h}_{\phi}(\omega_{1},\omega_{2};\tau)  =   \lim_{t\to \infty} \frac{\langle: \hat{A}^{\dagger}_{\omega_{1}}(t)  \hat{A}_{\omega_{1}}(t) \hat{A}_{\phi; \omega_{2}}(t+\tau) : \rangle}{\langle \hat{A}^{\dagger}_{\omega_{1}}(t)\hat{A}_{\omega_{1}}(t) \rangle \langle \hat{A}_{\phi; \omega_{2}}(t+\tau) \rangle}, 
\label{eq:corrfreq1}
\end{equation}
where $\hat{A}_{\omega_{i}}(t)  =  \int_{-\infty}^{t} e^{(i\omega_{i}-\Gamma_{i}/2)(t-t_{1})}\hat{a}(t_{1})dt_{1}$ is construed as the field amplitude detected at frequency $\omega_{i}$, within the frequency window $\Gamma_{i}$, at time $t$ \cite{gerard}. The frequency windows would be taken to be Lorentzian filters placed at frequencies $\omega_{1/2}=\omega_{S/aS}$, where the subscript stands for Stokes/anti-Stokes locations. Notwithstanding the reasonableness of this proposal, its actual computation would represent itself a cumbersome task in its present integral form.  \\

To circumvent the problem of computing correlations of the type (\ref{eq:corrfreq1}), one can resort to the sensing method proposed by E. del Valle and co-workers \cite{elena} consisting in the inclusion of ancillary frequency-resolved sensors subsumed into the whole dynamics of the system we seek to describe. According to this proposal, each sensor can be viewed, for instance, as a two-level system with a  annihilation (creation) operator $\varsigma_{i}$ ($\varsigma_{i}^{\dagger}$) and a transition frequency $\omega_{i}$, whose life time is given by the inverse of the detector linewidth, $\Gamma_{i}^{-1}$. In order to prevent the dynamics of our central system from being altered by the sensors' presence, the strength of the coupling, $\varepsilon_{i}$, between the sensors and the corresponding spectral component of the field to be measured must fulfill the condition $\varepsilon_{i} \ll \sqrt{\Gamma_{i}\gamma_{Q}/2}$, where $\gamma_{Q}$ denotes the decay rate associated with the rung of the energy ladder involved in the corresponding transition; in our case, $\gamma_{Q}=\gamma_{m}$. Although Eq. (\ref{eq:corrfreq1}) is, strictly speaking, a third-order correlation function in the field amplitude, the deployment of only two sensors would be necessary for the desired correlation to be emulated: one of them registering the signal of the field emission, say, at frequency $\omega_{2}$, while the other one being intended for recording the intensity of the spectral component centered at $\omega_{1}$. In doing so, the master equation (\ref{eq:mastereq1}) has to be extended accordingly by adding $H_{s}=\sum_{i} [\omega_{i}\hat{\varsigma}_{i}^{\dagger}\hat{\varsigma}_{i}+\varepsilon_{i} (\hat{a} \hat{\varsigma}_{i}^{\dagger}+\hat{a}^{\dagger}\hat{\varsigma}_{i})]$ to the original Hamiltonian (\ref{eq:ham2}), a dipole-like interaction term, reminiscent of that of the standard Jaynes-Cummings model, and the decay terms $\sum_{i} \frac{\Gamma_{i}}{2} \mathcal{L}_{\hat{\varsigma}_{i}}[\hat{\rho}]$ to the dissipative part; for the sake of simplicity, the $\Gamma_{i}$'s are chosen to match the widths of the Stokes spectral components, the latter being approximately given by $\gamma_{m}$ \cite{javier1}. Since the structure of the aforesaid interaction would make us think of each sensor as playing the role of a two-level atom dipolarly and very weakly coupled the field, it is surmised that, on the basis of this analogy, the sought correlation between sensors may be deployed as
\begin{equation}
h_{\phi}(\omega_{1},\omega_{2},\tau)  =  \lim_{\varepsilon_{1}, \varepsilon_{2} \to 0} \frac{\langle : \hat{\varsigma}_{1}^{\dagger}(0) \hat{\varsigma}_{1}(0) \hat{\varsigma}_{\phi;2} (\tau) : \rangle_{ss}}{\langle \hat{\varsigma}_{1}^{\dagger}(0) \hat{\varsigma}_{1}(0)\rangle_{ss} \langle \hat{\varsigma}_{\phi;2}(0)\rangle_{ss}},
\label{eq:hext}
\end{equation}
with $\hat{\varsigma}_{\phi;2}=(\hat{\varsigma}_{2} e^{i\phi}+\hat{\varsigma}_{2}^{\dagger}e^{-i\phi})/2$. Thus, the sensed information is now directly extracted from the dynamical variables of the two two-state  systems; the construction itself is feasible from the view point of the sensing method described in the supplemental material of Ref. \cite{elena}, and its structure is inspired by the work of Marquina-Cruz and Castro-Beltr\'an \cite{marquina}. Even though this criterion does not enable us to establish rigorously the possible equivalence between (\ref{eq:corrfreq1}) and (\ref{eq:hext}), the former being conjectural in itself, let us adopt the latter as an assay for exploring, as another complement to the present fact-finding examination, the simultaneous quadrature-photon detection, $\tau=0$, focusing on a given quadrature, say, 
\begin{equation}
h_{0}(\omega_{1},\omega_{2},0)  =   \lim_{\varepsilon_{1}, \varepsilon_{2} \to 0} \frac{\langle  \hat{n}_{1/2}(0) \hat{\varsigma}_{0;2/1}(0) \rangle_{ss}}{\langle \hat{n}_{1/2}(0) \rangle_{ss} \langle \hat{\varsigma}_{0;2/1}(0)\rangle_{ss}}, 
\label{eq:hsim}
\end{equation}
with the shorthand notation $\hat{n}_{1/2}=\hat{\varsigma}_{1/2}^{\dagger} \hat{\varsigma}_{1/2}$ and $\hat{\varsigma}_{0;2/1}=(\hat{\varsigma}_{2/1} +\hat{\varsigma}_{2/1}^{\dagger})/2$; the subindexes  highlight the interchangeability of the corresponding quantities. 
\begin{figure}[t!]
\includegraphics[width=9.5cm, height=8cm]{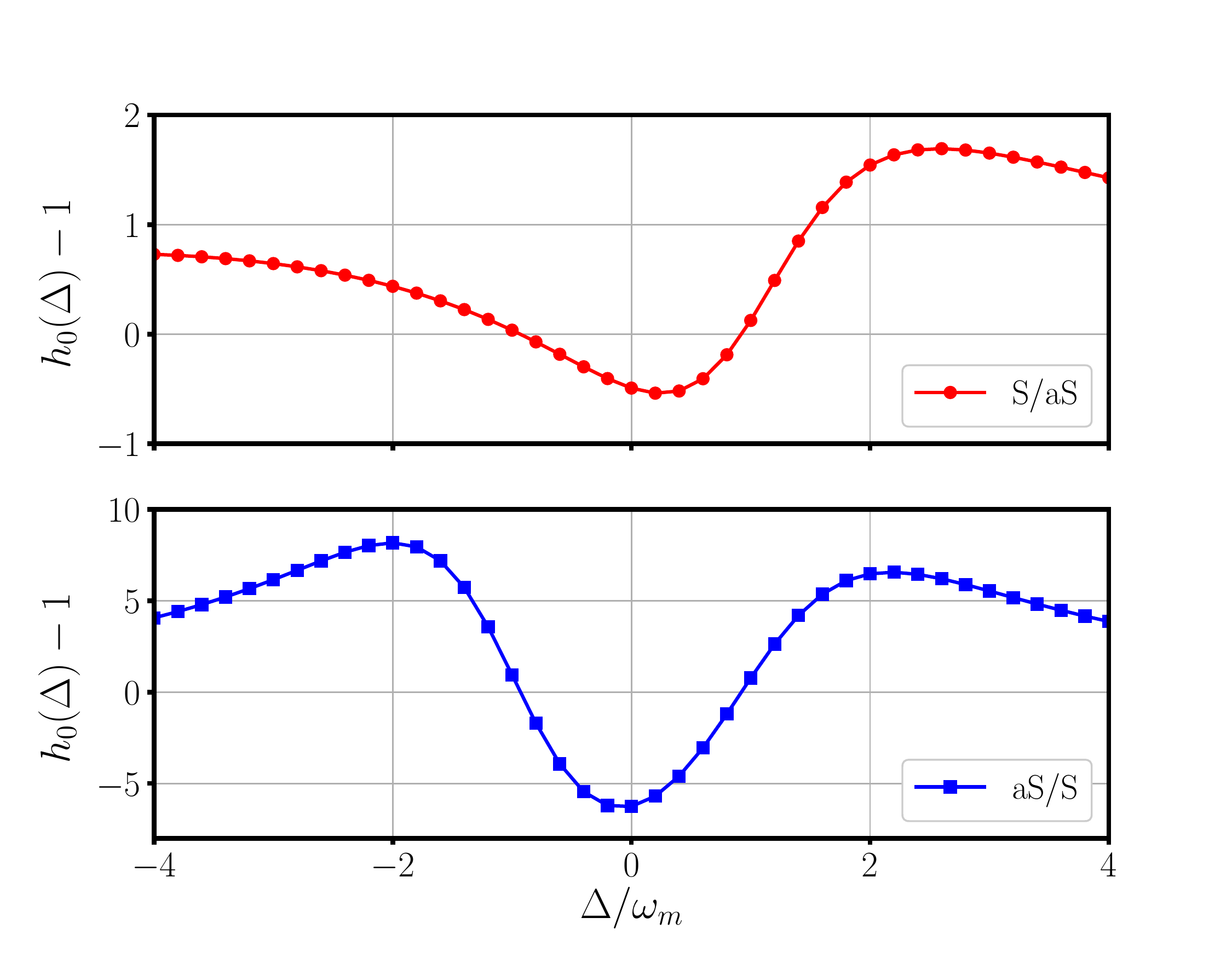} 
\caption{Frequency-filtered intensity-field correlations at zero delay, $\tau=0$, as a function of the scaled detuning $\Delta/\omega_{m}$, with frequencies of detection located at $\omega_{1/2}=\mp \omega_{m}$, sensor linewidths $\Gamma_{1/2} = \gamma_{m}$ and couplings $\varepsilon_{1/2} =10^{-4}\omega_{m}$. The parameters used are: $g=5$ meV, $\phi=0$ and $\Omega=0.3\omega_{m}$. The upper (lower) panel shows the correlation between the quadrature of the anti-Stokes (Stokes) signal and the intensity of the Stokes (anti-Stokes) one.}
\label{figure9}  
\end{figure}
We stress that this algebraic strategy would be tantamount to somehow adjusting the actual detectors in the CHD setup previously delineated so as to operate them under the frequency filtering performance, giving us the possibility to probe, at least, two undoubtedly accomplishable scenarios from the experimental point of view: i) one of them, labeled as the S/aS-event for shortness, is that in which the two detectors constituting the BHD arm, see Fig. \ref{figure3}, are in charge of registering the field amplitude corresponding to anti-Stokes photon emission, while the detector $D_{I}$ does the same for the intensity of the Stokes constituent; ii) the converse scenario, the aS/S-event, is such that the BHD port registers the quadrature of the Stokes signal, whereas the intensity of the anti-Stokes signal is measured by actual detector $D_{I}$. The outcome of the aforesaid S/aS- and aS/S-events provided by Eq. (\ref{eq:hsim}) is displayed, respectively, in the upper and lower panels of Fig. \ref{figure9} in keeping with the spirit of looking into the effect of heating ($\Delta <0$) or cooling ($\Delta >0$) of the molecule; the calculation is also restricted to the particular case of a very weak pump laser, $\Omega=0.3 \omega_{m}$. One sees that the frequency selectiveness accentuates the nonclassical features of the scattered photons in the light of the clear violation of both classical inequalities (\ref{eq:ineq1}) and (\ref{eq:ineq2}); this is particularly emphasized in the case of the aS/S-events (lower panel), regardless of whether the molecule is cooled or heated. In the S/aS events, on the other hand, the nonclassical features are the most marked when cooling the molecule (upper panel). \\   


\section{Concluding remarks}

On the basis of the algebraic framework provided by the quantum-mechanical model of SERS, some aspects of phase-dependent fluctuations of the light inelastically scattered by a single molecule placed inside the plasmon field of a cavity were explored, theoretically, through using the technique of conditional homodyne detection. Specifically, intensity-field correlations obtained by this technique made it possible to assess, in both time and frequency domains, the degree of non-Gaussianity and non-classicality of Raman photons mainly from the perspective of color-blind correlation measurements. Despite this unselective approach, given that we dealt with broadband detectors in the CHD setup in the first instance, it already reveals the extent to which the phase-dependent dynamics of the scattered light may be activated by the molecule's presence in the cavity. In this regard, the following issues are highlighted: 
\begin{itemize}
\item It is found that, by splitting the correlation functions into their second- and third-order fluctuation components, the origin of the non-classicality, for a moderate pump laser, is almost entirely due to the second-order fluctuations,  signaling squeezing within frequency domains located around the Stokes and anti-Stokes positions. Given this, the deviation from Gaussian noise is not significant.       
\item The heating or cooling of the molecular vibrations also led to discernible effects on the  correlation measurements, the former case yielding slightly larger fluctuations than latter one. However, for the out-of-phase quadrature of the field, even more conspicuous fluctuations are found in the case of zero detuning, as well as bolstering their time  asymmetry.  
\item Our discussions have also been briefly supplemented by putting forward the possibility to carry out frequency-filtered intensity-field correlations based on the same CHD set up. In the light of the outcome of an specific example concerning simultaneous intensity-field measurements, one can foresee the manifestation of high correlations between specific Raman photons, even if a very weak pump laser is taken into consideration. This fact, at variance with the color-blind scheme, heralds, in this context, highly nonclassical scattered light by virtue of the violation of the inequalities that take part in CHD (Eqs. (\ref{eq:ineq1}) and (\ref{eq:ineq1})). The feasibility of the proposed procedure for computing such  correlations remains, however, surmise and a theoretical method to cope with their computation is expected to be firmly established elsewhere.
\end{itemize}

Unlike standard homodyne systems, CHD is regarded, owing to its conditioning attribute, as a particularly convenient tool to investigate non-classical and non-Gaussian properties of the light source it seeks to describe that goes beyond analyzing the squeezing itself, even for very weak light sources. Although the partitioning of the corresponding correlation into second- and third-order fluctuations cannot be directly realizable from the experimental viewpoint, they represent valuable theoretical information that enables us to unveil the main contributions to the phase-dependent dynamics, non-Gaussianity and non-classicality of light in a given context-dependent circumstance. These advantages of using the CHD framework have already been highlighted in theoretical studies of two- and three-level atom resonance fluorescence \cite{castro1,castro3}. \\

Another issue that would deserve to be addressed is to investigate the influence of the anharmonicity of the molecule (viewed, for instance, as a Morse-like oscillator \cite{frank,carvajal}) on the outcome of either color-blind or frequency-filtered correlation measurements through the proper modification to the current optomechanical description of SERS; this more general situation will hopefully be treated elsewhere. The present results, for the time being, are expected to be a modest first step towards pointing out the importance of undertaking a deeper exploration and understanding on the role of phase-dependent correlations in single-molecule SERS studies, thus triggering the experimental and/or theoretical interest from the quantum optics and SERS communities \cite{mark}.


\section*{Acknowledgments}
I want to thank Prof. H. M. Castro-Beltr\'an for reading the manuscript. Thanks also to I. Ramos-Prieto for introducing me to the use of QuTip library in Python and for his help with Figs. \ref{figure1} and \ref{figure3}, and Prof. J. R\'ecamier for his kind hospitality at ICF.


\end{document}